\documentclass[twocolumn]{aastex63}

\pdfoutput=1

\usepackage{amssymb}
\usepackage{amsmath}
\usepackage{amsfonts}
\usepackage{graphicx}
\usepackage{color}
\usepackage{hyperref}
\usepackage[caption=false]{subfig}
\usepackage{mathtools}
\usepackage[normalem]{ulem}

\def\lsim{~\rlap{$<$}{\lower 1.0ex\hbox{$\sim$}}}
\def\bsim{~\rlap{$>$}{\lower 1.0ex\hbox{$\sim$}}}

\def\hmpc{\ {\rm {\it h}^{-1}Mpc}}

\def\hmsun{\ {\rm {\it h}}^{-1}M_\odot}

\def\eV{\ {\rm eV}}

\def\Kel{\ {\rm K}}

\def\kpc{\ {\rm kpc}}
\def\ergss{\ {\rm erg\, s^{-1}}}

\def\apjl{Astrophys.\ J.\ Lett.}

\def\aap{Astron.\ Astrophys.}

\def\apjs{Astrophys.\ J. Supp.}

\newcommand{\Halpha}{\mathrm{H}\alpha}

\newcommand{\cHII}{C_{\rm{HII}}}

\allowdisplaybreaks

\usepackage[normalem]{ulem}

\usepackage{lineno} % to add numbers
\begin{document}
%\linenumbers

\title{Constraining Baryonic Feedback at $z\sim 1$ with the $\Halpha$ Luminosity Function}

\author[0009-0007-7887-783X]{Ivan Rapoport}%\,\orcidlink{0009-0007-7887-783X}
\email{ivanr@campus.technion.ac.il}
\affiliation{Physics Department, Technion -- Israel Institute of Technology, Haifa 3200003, Israel}

\author[0000-0003-2062-8172]{Vincent Desjacques}
%\email{dvince@physics.technion.ac.il}
\affiliation{Physics Department, Technion -- Israel Institute of Technology, Haifa 3200003, Israel}

\author[0000-0001-9735-4873]{Ehud Behar}
%\email{behar@physics.technion.ac.il}
\affiliation{Physics Department, Technion -- Israel Institute of Technology, Haifa 3200003, Israel}

\author[0000-0002-0404-003X]{Shmuel Bialy}
%\email{behar@physics.technion.ac.il}
\affiliation{Physics Department, Technion -- Israel Institute of Technology, Haifa 3200003, Israel}

\date{\today}

\begin{abstract}
Baryonic feedback remains one of the primary uncertainties in galaxy formation models because of its impact on the interstellar medium (ISM) and star formation. We investigate its effect on the $\Halpha$ luminosity function (LF) using the Illustris and IllustrisTNG simulations. We incorporate direction- and metallicity-dependent dust attenuation, thermal sputtering and a physically motivated treatment of the propagation of Lyman-series photons. We model three sources of $\Halpha$ emission: star-forming H II regions, collisional excitation and recombination in the diffuse ISM, and AGN-induced photo-excitation of diffuse gas. The models are calibrated against observed $\Halpha$ LF at $z\sim1-1.5$ using a single free parameter relating the star formation rate to the H II-region $\Halpha$ luminosity. We find that, for $L_{\Halpha}\geq 10^{41}\ergss$, emission from diffuse gas contributes $\sim 5\%-15\%$ of the total $\Halpha$ luminosity in both simulation sets. AGN photo-excitation produces only a modest enhancement of the bright end of the LF, although it dominates the $\Halpha$ emission of Illustris galaxies with $L_{\Halpha}\gtrsim 10^{43}\ergss$. The low-feedback IllustrisTNG simulations provide a good fit to the observed LF for all emission models considered, whereas the high-feedback Illustris predictions are difficult to reconcile with both the bright and faint end of the LF. This tension between $\Halpha$ LF and stacked kSZ measurements highlights the combined power of these probes for constraining the interplay between baryonic feedback, gas properties, and the physics governing emission-line galaxies.
\end{abstract}

%----------------------------------------------------------------------------%
\section{Introduction}
\label{sec:intro}
%----------------------------------------------------------------------------%
Emission-line galaxies (ELGs) are among the most important tracers of the large-scale structure of the Universe. Their characteristic strong emission lines enable efficient spectroscopic redshift measurements across a wide range of redshifts, making them prime targets for modern cosmological surveys such as \textit{Euclid}, DESI, SPHEREx and NGRST \citep{euclidcollaboration2024,desicollaboration2016,spherexcollaboration2014,spergel2015}. Large samples of ELGs have therefore become a cornerstone of efforts to map the cosmic web and constrain the growth of structure, dark energy, and the underlying cosmological model \citep[e.g.][]{peebles:1980,kaiser:1987,efstathiou/etal:1990,ballinger/peacock/heavens:1996,peacock/etal:2001,tegmark/etal:2004,cole/etal:2005,eisenstein/etal:2005,guzzo/etal:2008,anderson/etal:2012,sdssiv,demattia/etal:2021,zhao/etal:2021,ivanov:2021,DESI2024iii,DESI2024v,euclidcollaboration2024,chudaykin/etal:2026}.

Among the various emission lines observed in galaxies, the $\Halpha$ 6564$\text{\AA}$ line is one of the most widely used probes of recent star formation and the gas component of galaxies \citep{Kennicutt_1983, Calzetti2012}, with $\Halpha$ luminosity from H II regions commonly related to the star formation rate through an empirically calibrated conversion factor (denoted $\cHII$ below). $\Halpha$ emission arises through multiple physical channels within the interstellar medium (ISM), most importantly photoionization of neutral hydrogen by short-lived massive stars followed by recombination, but also through electron impact excitation and recombination in the warm and hot phases of the ISM, as well as photo-ionization by active galactic nuclei (AGN). Consequently, $\Halpha$ production is closely coupled to the evolution of stellar populations, AGN activity, and the thermodynamic state of the gas \citep[e.g.][]{Tacchella_2022}. Despite the importance of $\Halpha$ observations, however, the relative contributions of these different emission mechanisms remain poorly constrained \citep[e.g.][]{Hu_2024}, since the relative importance of each channel depends on the ISM's density and thermal state, which are in turn governed by star formation and feedback processes.

The $\Halpha$ luminosity function (LF) is a fundamental observable for studies of ELG evolution. The close connection between $\Halpha$ emission and star formation has made the $\Halpha$ LF a widely used probe of the cosmic star formation rate density and its evolution with redshift \citep{tresse/etal:2002, villar/etal:2008, sobral/etal:2009, sobral/etal:2011, Lee/etal:2012, kennicutt/evans:2012, ibar/etal:2013, coughlin/etal:2018, boselli/etal:2023}. The increasing precision and redshift coverage of $\Halpha$ surveys have therefore provided important constraints on the build-up of stellar mass across cosmic time. Interpreting these observations, however, requires accurate modeling of the physical processes that govern the production and propagation of $\Halpha$ photons within galaxies, including dust attenuation in the ISM of the source galaxies. The absorption and scattering of $\Halpha$ photons along the line of sight attenuate the observed emission, affecting thereby predictions of $\Halpha$ LF and the clustering properties of ELGs \citep{Hayes_2010, Buat_2018}.

Recent analyses of kinetic Sunyaev-Zel'dovich (kSZ) observations using the Illustris and IllustrisTNG simulations have shown that the stacked kSZ signal is better reproduced by the high Illustris feedback model \citep{Hadzhyska_2025, Guachalla_2025, bigwood/etal:2025}. 
Here, "high feedback" refers to the boosted black-hole accretion and the radio-mode outflows in Illustris, which are curtailed in IllustrisTNG.
In the present study, we use the same simulations to test which feedback scenario best reproduces the observed $\Halpha$ LF at redshifts $z\sim 1$ and $1.5$.
Testing this scenario against $\Halpha$ data, however, requires a more accurate treatment of the processes governing $\Halpha$ production and escape than has previously been used. In particular, the model of \citet{Rapoport_2025} suffers from three main limitations. First, its treatment of star-forming cells assumes a fixed fraction of the cell's mass to reside in the hot phase and relies on a phase-averaged temperature that does not correspond to a physical gas state. Second, its treatment of AGN radiation attenuation assumes an a-priori attenuation scale rather than one derived from the actual distribution of neutral hydrogen in simulated galaxies. Third, its treatment of dust attenuation applies a single, fixed $\Halpha$ extinction value across all galaxies, irrespective of their individual properties. For this purpose, we improve the $\Halpha$ emission line model presented in \citet{Rapoport_2025} by including (i) a refined treatment of the star-forming cells, (ii) a physically-motivated computation of dust extinction, (iii) a refined computation of the $\Halpha$ emission induced by AGN radiation, and (iv) a physically motivated model for the H\,II region conversion parameter $\cHII$, for which we derive an expected value.

This paper is organized as follows. In \S\ref{sec:illustrisTNG}, we introduce the simulations used in this work. In \S\ref{sec:emission_model}, we summarize the $\Halpha$ emission model and describe the improvements introduced here. In \S\ref{sec:dust}, we present our galaxy-by-galaxy calculation of dust extinction. In \S\ref{sec:fit}, we fit three variations of the $\Halpha$ emission model to observed LF data $z\sim1-1.5$. We discuss our results in \S\ref{sec:discussion} and summarize our conclusions in \S\ref{sec:conclusions}.

%----------------------------------------------------------------------------%
\section{The Illustris-TNG simulations}
\label{sec:illustrisTNG}
%----------------------------------------------------------------------------%
The IllustrisTNG project comprises a suite of large-scale cosmological simulations that model galaxy formation using gravity and magnetohydrodynamics \citep{TNG_DR, TNG_1, TNG_2, TNG_3, TNG_4, TNG_5, TNG_6, TNG_7}. The Illustris and IllustrisTNG simulations adopt slightly different $\Lambda$CDM parameter sets (based on the results of \citet{WMAP9} and \citet{Planck2015}, respectively), which are broadly consistent within observational uncertainties. In this work, we analyze the TNG100-1 simulation (TNG, for short) alongside the earlier Illustris-1 run. These simulations have a similar resolution, with a box length of $75\hmpc$ and an average gas cell mass of approximately $10^6\hmsun$. To assess the impact of resolution on our modeling, we also examine the larger-volume TNG300-1 simulation, which has a box length of $205\hmpc$ and an average gas cell mass of $\approx7.5\times 10^{6}\hmsun$.

In addition to differences in cosmological parameters, TNG employs an updated baryonic feedback framework, which incorporates a revised treatment of black hole growth and galactic winds, the inclusion of magnetic fields, and a dual-mode AGN feedback. These changes mitigate a number of discrepancies between the Illustris simulations and benchmark observations, resulting in more realistic galaxy sizes and morphological distributions \citep{TNG_7}. The TNG model reproduces a wide range of observational constraints pertaining to galaxy formation. On galactic scales, it broadly matches the cosmic star formation rate density and the star-forming main sequence, as well as the observed galaxy stellar mass function and its redshift evolution \citep{Donnari2019, TNG_7}. It also reproduces the mass-metallicity relation at low redshift for galaxies with stellar mass in the range $10^9\lesssim M_*/M_\odot \lesssim 10^{10.5}$ \citep{Torrey_2019}, which is relevant for the dust modeling presented in \S\ref{sec:dust}, and matches baryon fraction constraints for massive clusters \citep{Artale_2021}.

%----------------------------------------------------------------------------%
\section{$\Halpha$ emission modelling}
\label{sec:emission_model}

We adopt the framework introduced in \cite{Rapoport_2025} and post-process the outputs of hydrodynamical simulations of galaxy formation to model $\Halpha$ emitters. 
This atomic-physics based model of $\Halpha$ emission takes into account multiple emission channels evaluated at the level of individual gas cells, 
which allows for the construction of spatially resolved $\Halpha$ emission maps. 
The computation of the $\Halpha$ emission assumes steady-state populations of hydrogen atomic levels and is carried out within the coronal approximation. 
This second assumption implies that excitations proceed either from the ground state of neutral hydrogen or through (radiative) recombination of ionized hydrogen. 
These simplifications are appropriate for the optically thin interstellar medium. A concise summary of the implementation adopted here is provided in \S\ref{sec:appendix}; we refer the reader to \citet{Rapoport_2025} for a complete description.

\subsection{Physical mechanisms for $\Halpha$ emission}

The original model of \cite{Rapoport_2025} includes $\Halpha$ photons produced by spontaneous radiative decay of the electronic $n=3$ level of atomic hydrogen into $n=2$, where the $n=3$ level is populated by several excitation processes acting on neutral hydrogen in the diffuse ISM: collisional excitation (CE), photo-excitation by stellar radiation, and photo-excitation by AGN radiation. We additionally include a contribution from radiative recombination (RR) of ionized hydrogen, and we separately model the effective contribution from unresolved H II regions. In this work, we omit the stellar photo-excitation contribution to the diffuse ISM because, like the H II-region contribution, its amplitude scales with the local star formation rate and therefore does not introduce an independent degree of freedom. Excluding this term simplifies the model without significantly affecting the predicted $\Halpha$ luminosities.

Consequently, we can write the total intrinsic $\Halpha$ luminosity of a galaxy as the following sum of emission components:
\begin{equation}
    \label{eq:totalHaLum_intr}
    L_{\Halpha}^{\text{intr}} =L_{\Halpha}^{\text{CE}}+L_{\Halpha}^{\text{RR}}+L_{\Halpha}^{\text{AGN}}+L_{\Halpha}^{\text{HII}}\;.
\end{equation}
Where $L_{\Halpha}^{\text{CE}},\ L_{\Halpha}^{\text{RR}},\ L_{\Halpha}^{\text{AGN}}$ and $L_{\Halpha}^{\text{HII}}$ are the intrinsic galactic luminosities of the aforementioned emission components. Here, each term is computed by summing the relevant contribution over all gas cells belonging to the host subhalo. 
In addition to the local thermodynamic properties and ionization state provided by the simulations, the $\Halpha$ emission depends on the following global parameters: 
\begin{itemize}
\item The slope of the AGN spectral energy distribution (SED), $\alpha_{\text{AGN}}$, which determines the rate of $\Halpha$ photon production (in $\rm s^{-1}$) per H\,I atom in a given cell with incident AGN bolometric flux $f_{\rm{AGN}}$ (see \S\ref{sec:AGN}).
\item The conversion parameter $\cHII$ (in $\ergss \text{yr}M_\odot^{-1}$), which determines the effective $\Halpha$ luminosity arising from H II regions according to
\begin{equation}
    L_{\Halpha, i}^{\text{HII}} = \cHII \text{SFR}_i  \;,
\end{equation}
where $\text{SFR}_i$ (in $M_\odot \text{yr}^{-1}$) is the star formation rate of the $i$th cell.
\end{itemize}
Following \cite{Rapoport_2025}, we set $\alpha_{\text{AGN}}=2$. By contrast, $\cHII$ is treated as a free parameter and constrained by fitting the model predictions to data.

The IllustrisTNG simulations employ the effective two-phase ISM model of \citet{Springel_2003}, in which the thermodynamic properties of star-forming cells represent averages over unresolved cold and hot phases. Unlike \citet{Rapoport_2025}, which assumed a fixed hot-phase mass fraction, we recover the hot and cold components by inverting the effective internal energy following the \citet{Springel_2003} model, as described in \S\ref{sec:appendix}. In this framework, the hot-phase mass fraction is $\sim10\%$ near the star-formation threshold and decreases with increasing density. We apply our atomic emission model only to the hot phase, since the cold phase is, by construction, the star-forming gas already accounted for by the H II-region component, and including it here would double-count that contribution.

\subsection{AGN-induced $\Halpha$ emission}
\label{sec:AGN}

To model the $\Halpha$ emission arising from photo-excitation of neutral hydrogen by Lyman-series photons emitted by AGN, we follow \citet{Rapoport_2025} and treat the AGN 
as a point source located at the galactic center\footnote{The \texttt{SubhaloPos} field.}, with a power-law SED characterized by a fixed slope $\alpha_\mathrm{AGN}=2$. 

The primary modification introduced in this work concerns the attenuation of the AGN radiation field away from the center. In \citet{Rapoport_2025}, the AGN flux incident on each cell was attenuated by a 
transmission function $\mathcal{T}(r_i)=\exp{\left(-r_i/\lambda_{\rm att}\right)}$, where $r_i$ is the distance of the cell from the galactic center and $\lambda_{\rm att}=1\kpc$ is a fixed attenuation length representing the effective removal of resonant photons from the radiation field. In this prescription, $\lambda_{\rm att}$ is assumed a priori and is independent of the neutral hydrogen distribution within a given galaxy. In this work, we calculate instead the transmission $\mathcal{T}(r_i)$ through the simulated gas distribution on a cell-by-cell basis, taking into account the positions and H\,I content of individual gas cells along the path from the AGN to each cell receiving its radiation.

The attenuation considered here arises from the resonant absorption of Lyman lines by ISM hydrogen. Following absorption, an excited atom may either re-emit a photon in the same Lyman transition or undergo a radiative cascade through intermediate levels. In the latter case, the absorbed energy is redistributed among multiple lower-energy photons, of which most do not match the resonance energies for exciting H\,I from the ground state. As a result, the population of photo-exciting Lyman photons decreases as the radiation propagates through the galaxy. We model this net loss using an effective extinction cross section, $\sigma_{\rm H\,I,ext}^{\rm Ly}$. We do not separately account for dust extinction of these Lyman-series photons, since the dust cross section is negligible compared to the resonant H\,I interactions considered here. To construct the transmission function we first estimate $\sigma_{\rm H\,I,ext}^{\rm Ly}$, in the coronal approximation this cross section can be computed as
\begin{equation}
    \sigma_{\text{H\,I,ext}}^{\text{Ly}}\approx \sum_{j} \sigma_{1j}^{\text{PE}}\left(1-\frac{A_{j1}}{\sum_k A_{jk}}\right) \;,
\end{equation}
where $\sigma_{1j}^{\text{PE}}$ is the photo-excitation cross section for the $1\to j$ transition, and $A_{jk}$ are the Einstein coefficients for radiative decay from level $j$ to level $k$. The factor in parentheses gives the probability that the excited atom does \textbf{not} relax via the $j \to 1$ channel.

We adopt line-center cross sections assuming purely thermal-broadened profiles, valid since the natural linewidth of Lyman series lines is negligible compared to the Doppler width at ISM temperatures: 
\begin{equation}
    \sigma_{1j}^{\text{PE}} = \frac{he^2}{m_ec}f_{1j} \frac{1}{E_{1j}}\sqrt{\frac{\pi m_e c^2}{2kT_i}} \;,
\end{equation}
where $h$ is Planck’s constant; $e$ and $m_e$ are the electron charge and mass; $c$ is the speed of light; $f_{1j}$ and $E_{1j}$ are the oscillator strength and transition energy for the $1\to j$ transition; $k$ is the Boltzmann constant; and $T_i$ is hot-phase temperature of the cell. Line-center values represent an upper limit on $\sigma_{1j}^{\text{PE}}$; consequently, our computed transmissions $\mathcal{T}(r_i)$ underestimate the true transmission, and the AGN-induced $\Halpha$ contribution could be somewhat higher than what we report below. 
Using the \texttt{CHIANTI} atomic database \citep{Chianti_1, Chianti_2}, we obtain the energies and oscillator strengths for all available hydrogen transitions and find
\begin{equation}
    \sigma_{\text{H I,ext}}^{\text{Ly}} \approx 4.49\times 10^{-17} \left(\frac{10^4 \Kel}{T_i}\right)^{1/2}\ \rm{cm^2/H\,I}
\end{equation}
at line center. The Lyman-series transmission to a gas cell $i$ is computed by tracing the line of sight from the AGN to the cell and summing the contributions from all intersecting gas cells along the ray:
\begin{equation}
    \mathcal{T}(r_i)=\exp\left[-\sum_{j\in\mathcal{R}_i} \sigma_{\text{H I,ext}}^{\text{Ly}}(T_j) n_{\rm H\,I,j} \Delta s_j \right]
\end{equation}
where $\mathcal{R}_i$ denotes the set of gas cells intersected by the ray to cell $i$, $n_{\rm H\,I,j}$ is the neutral hydrogen number density of cell $j$, and $\Delta s_j$ is the ray intersection path length within that cell.

To reduce the computational cost, we neglect the attenuation calculation for cells at radii $r_i > 0.25 r_{1/2}$, setting $\mathcal{T}(r_i)=0$ beyond this limit, where $r_{1/2}$ is the galactic gas half-mass radius. At this radius, the mean optical depth to Lyman-series photons is already $\approx 2$, so the AGN-induced emission beyond it is strongly suppressed.

%----------------
\begin{figure}
    \centering
    \includegraphics[width=0.42\textwidth]{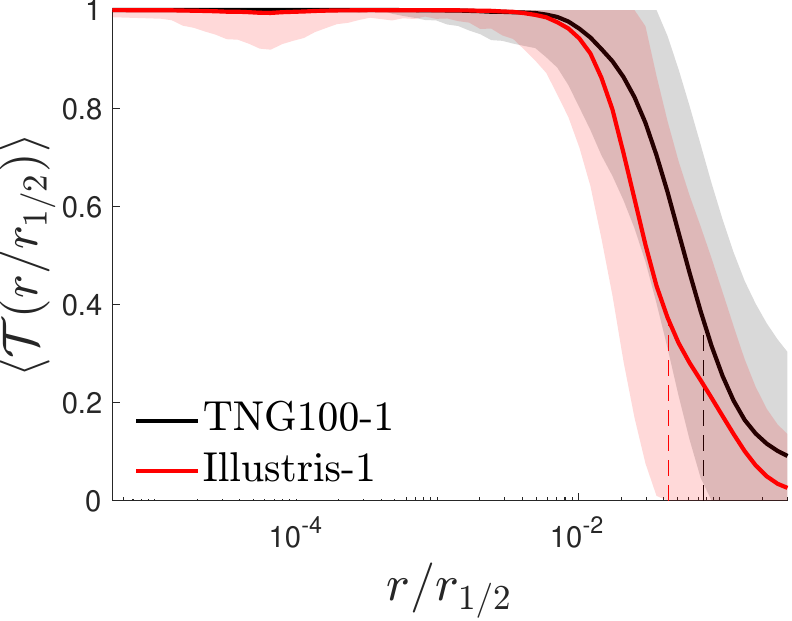}
    \caption{Stacked angularly-averaged transmission profiles of Lyman-series photons emitted from the galactic center ($r=0$) for AGN-host galaxies in the two primary simulations at $z=1$, shown as a function of the scaled radius $r/r_{1/2}$. Solid lines denote the mean transmission profile, while the shaded regions indicate the $1\sigma$ scatter among galaxies. The vertical dashed lines mark the radius $r/r_{1/2} = 0.04-0.07$ at which the average optical depth already reaches unity, corresponding to $\ln \mathcal{\langle T \rangle }=-1$.}
    \label{fig:AGN_transmission}
\end{figure}
%--------------
In Fig.~\ref{fig:AGN_transmission} we show the stacked transmission profile $\langle\mathcal{T}(r/r_{1/2})\rangle$, obtained by averaging the radial transmission profiles of all AGN-host galaxies\footnote{Galaxies with a non-zero black hole accretion rate, the \texttt{SubhaloBHMdot} field.} in bins of scaled radius $r/r_{1/2}$. 
Results are shown for the two primary simulations at $z=1$. Differences in the feedback prescriptions of different simulations lead to distinct H\,I distributions in the central regions of galaxies and, consequently, to different transmission profiles. 

The average optical depth reaches unity at $r/r_{1/2}\approx 0.04$ and $0.07$ for Illustris and TNG, respectively. Adopting the mean $r_{1/2}$ of AGN-host galaxies in each simulation, these values 
correspond to physical distances of approximately $0.9$ and $2.4\kpc$. This indicates that the attenuation of Lyman-series photons—and therefore the resulting AGN-induced $\Halpha$ emission—is governed primarily by the thermodynamic state of the gas within the central few kpc of the galaxy.

\section{Dust absorption and attenuation}
\label{sec:dust}
Our refined $\Halpha$ line model includes a more comprehensive treatment of the impact of interstellar dust than in \citet{Rapoport_2025}, where a uniform extinction was assumed for all galaxies. In this section, we quantify the attenuation of $\Halpha$ photons along the line of sight as they escape the galaxy.

To compute the spatially and directionally resolved $\Halpha$ dust extinction of IllustrisTNG galaxies, we use for each gas cell the hydrogen number density $n_{\mathrm{H},i}$, cell volume $V_i=m_i/\rho_i$ (computed from the cell's total mass $m_i$ and density $\rho_i$), physical coordinates $\mathbf{x}_i$ and metallicity $Z_i$ (in solar units) provided by the simulations, as well as the cell's intrinsic $\Halpha$ luminosity $L_{\Halpha,i}^{\text{intr}}$. We select a single line-of-sight (LOS) direction, which is applied to all galaxies in the simulation. Our fiducial choice is the $z$ axis of the simulation box. Since galaxies are randomly oriented with respect to the fixed axes of the simulation box, a single direction samples the full range of inclinations across the galaxy population, so that population statistics such as the LF are unbiased by the choice of LOS. The attenuation of an individual galaxy, however, depends on its inclination. For convenience, we rotate the Cartesian coordinates of the simulations such that the new $z$-axis coincides with the LOS direction. In this new frame, the coordinates $(x,y)$ denote directions transverse to the LOS, while $z$ measures distance along the LOS away from the observer, which is taken to be located at $z_{\text{obs}}=-\infty$. All the geometric operations described below are performed in this LOS-oriented coordinate system.

For each gas cell $i$, we compute the hydrogen column density integrated along the LOS from the emission point toward the observer, where each gas cell is approximated as a uniform-density sphere with no internal structure. Since the simulation's gas cells form an unstructured, moving mesh rather than a regular grid, this requires explicitly determining which cells intersect a given LOS and their contributing path length. For $\Halpha$ photons emitted by cell $i$, we identify foreground gas cells $j$ that can contribute to $\Halpha$ extinction along the LOS. A cell $j$ is considered a candidate contributor if its perpendicular distance to the ray satisfies
\begin{equation}
    d_{\perp, j} = \sqrt{(x_j-x_i)^2 + (y_j-y_i)^2} < \left(\frac{3V_j}{4\pi}\right)^{1/3} \;.
\end{equation}
For each candidate cell, the half-chord length of the LOS through the spherical volume is
\begin{equation}
    \ell_{j,\text{half}}=\sqrt{\left(\frac{3V_j}{4\pi}\right)^{2/3} - d_{\perp, j}^2} \;,
\end{equation}
and the physical path length of the LOS inside cell $j$, truncated at the position of the emitting cell $i$, is then given by
\begin{align}
    & \Delta s_j = \max \left[0, \min\left(z_j + \ell_{j,\text{half}} ,\ z_i \right) - (z_j - \ell_{j,\text{half}}) \right] .
\end{align}
This expression correctly vanishes for cells lying entirely behind the emitting cell, reduces to the full chord length $2\,\ell_{j,\text{half}}$ for cells lying entirely in front, and interpolates between the two for intermediate configurations. The $\Halpha$ optical depth due to dust extinction experienced by photons leaving cell $i$ is thus given by
\begin{equation}
    \tau_{\Halpha, i} = \sum_{j\ \text{contributes\ to\ }i}  \sigma_{\text{ext}, j}^{\Halpha} n_{\text{H}, j} \Delta s_j \;.
\end{equation}

Following \citet{Remy_Ruyer_2014,Bialy_Sternberg_2019}, we scale the extinction cross section with the local dust-to-gas ratio, using each cell's metallicity $Z_j$ as a proxy:
\begin{equation}
    \label{eq:Kroupa_IMF}
      \sigma_{\text{ext}, j}^{\Halpha}(Z_j) = \sigma_{\text{ext}}^{\Halpha,\odot}\times\left\{
    \begin{array}{cl}
    Z_j & ,\ Z_j\geq Z_{\text{k}} \\
    Z_{\text{k}}\left(\frac{Z_j}{Z_{\text{k}}}\right)^{\alpha_{\text{DGR}}} & ,\ Z_j < Z_{\text{k}}
    \end{array} \right.
\end{equation}
where $\sigma_{\text{ext}}^{\Halpha,\odot}=4.07\times 10^{-22}\ \rm{cm^2/H}$ is the solar-metallicity $\Halpha$ extinction cross section \citep{Draine2011}, $Z_{\text{k}}=0.2$ and $\alpha_{\text{DGR}}=3$. This parametrization effectively makes low metallicity cells $(Z_j\lesssim Z_{\text{k}})$ transparent to $\Halpha$ radiation. These usually correspond to dilute cells that did not experience significant star formation. =Additionally, we account for the destruction of dust grains by thermal sputtering, where impacts by energetic gas particles erode dust grains at temperatures $T\gtrsim 10^5\Kel$ \citep[e.g.][]{Draine_Salpeter_1979b,McKinnon_2017}. Following a simple prescription, we set $\sigma_{\text{ext}, j}^{\Halpha}=0$ for cells with $T\geq 2\times 10^5\Kel$ to produce dust absorption values that are broadly consistent with the literature. This procedure is repeated for all gas cells $i$ within each galaxy to produce the total observed $\Halpha$ luminosity
\begin{equation}
\label{eq:totalHaLum}
    L_{\Halpha}=\sum_i L_{\Halpha,i}^{\text{intr}}\, e^{-\tau_{\Halpha,i}} \;.
\end{equation}

%----------------------------------------------------------------------------%
\section{Fitting luminosity function data}
\label{sec:fit}
%----------------------------------------------------------------------------%

\subsection{Best-fit models}

%----------------------------------
\begin{table*}
\centering
\hspace{-1.5cm}
\begin{tabular}{|c|c|c|c|c|}
\hline
\textbf{Simulations} & $z$ & $N_{\rm{sub}}$ & $N_{\rm{AGN}}$ & $N_{41}$ \\ 
& & & & \\
\hline
$\text{TNG100-1}$ & 1.0 & 327921 & 35190 & 13578  \\ \hline
$\text{TNG100-1}$ & 1.5 & 436218 & 33700 & 13692  \\ \hline
$\text{TNG300-1}$ & 1.0 & 2252850 & 680780 & 241416 \\ \hline
$\text{Illustris-1}$ & 1.0 & 357250 & 30315 & 14394 \\ \hline
$\text{Illustris-1}$ & 1.47 & 456033 & 28563 & 15961 \\ \hline

\end{tabular}
\caption{Number counts of simulated galaxy samples. For each simulation and snapshot redshift, the columns list the number of subhalos in the sample (subhalos with at least one gas cell), the number of subhalos hosting an AGN, and the number of ELGs with $L_{\Halpha}>10^{41}\ergss$ for the best-fit $\cHII$ in the HII+Coll+AGN model (see \S\ref{sec:fit}).}
\label{tab:subhalo_counts}
\end{table*}
%----------------------------------
\begin{table*}
\centering
\hspace{-1.5cm}
\begin{tabular}{|c|c|c|c|c|}
\hline
\textbf{Simulations} & $z$ & \rm{model} & $\log \cHII$ & $ \chi^2$ \\ 
& & & [$\rm{erg\ s^{-1}\ yr} \it M_\odot^{-1}$] & \\
\hline
$\text{TNG100-1}$ & 1.0 & HII & 41.335 & 0.7900 \\ \hline
$\text{TNG100-1}$ & 1.0 & HII+Coll & 41.295  & 0.9068 \\ \hline
$\text{TNG100-1}$ & 1.0 & HII+Coll+AGN & 41.295 & 0.9074 \\ \hline
$\text{TNG100-1}$ & 1.5 & HII & 41.22 & 0.7222 \\ \hline
$\text{TNG100-1}$ & 1.5 & HII+Coll & 41.16  & 0.5652 \\ \hline
$\text{TNG100-1}$ & 1.5 & HII+Coll+AGN & 41.16 & 0.5669 \\ \hline
$\text{TNG300-1}$ & 1.0 & HII & 41.62 & 0.7679 \\ \hline
$\text{TNG300-1}$ & 1.0 & HII+Coll & 41.59 & 0.7580 \\ \hline
$\text{TNG300-1}$ & 1.0 & HII+Coll+AGN & 41.585 & 0.7578 \\ \hline
$\text{Illustris-1}$ & 1.0 & HII & 41.29 & 2.0734 \\ \hline
$\text{Illustris-1}$ & 1.0 & HII+Coll & 41.25 & 2.2413 \\ \hline
$\text{Illustris-1}$ & 1.0 & HII+Coll+AGN & 41.25  & 2.2181 \\ \hline
$\text{Illustris-1}$ & 1.47 & HII & 41.19 & 4.0247 \\ \hline
$\text{Illustris-1}$ & 1.47 & HII+Coll & 41.16 & 3.5459 \\ \hline
$\text{Illustris-1}$ & 1.47 & HII+Coll+AGN & 41.16 & 3.6865 \\ \hline
\end{tabular}
\caption{Best-fit parameters for the different examined simulations and models. The columns show, in order: the snapshot redshift, the $\Halpha$ emission model, the best-fit $\log \cHII$ of the model and the normalized $\chi^2$ of the best fit.}
\label{tab:best_fit_properties}
\end{table*}
%----------------------------------

We consider three variations of our $\Halpha$ emission line model:
\begin{itemize}
    \item HII, where the H II regions solely contribute to the $\Halpha$ emission: $L_{\Halpha}=\sum_i L_{\Halpha,i}^{\text{HII}}e^{-\tau_{\Halpha,i}}$.
    \item HII+Coll, which also incorporates the collisional components: $L_{\Halpha}=\sum_i (L_{\Halpha,i}^{\text{HII}}+L_{\Halpha,i}^{\text{CE}}+L_{\Halpha,i}^{\text{RR}})e^{-\tau_{\Halpha,i}}$.
    \item HII+Coll+AGN, which includes all the sources of $\Halpha$ emission quoted in Eq.~\ref{eq:totalHaLum}. 
\end{itemize}
Note that the $\Halpha$ emission from the Coll component is parameter-free and therefore non-adjustable, that is, it is always present by construction.
Nevertheless, we have considered a HII model where Coll is artificially turned off in order to assess whether emission from H II regions alone is able to reproduce the data.

With $\alpha_\text{AGN}=2$ fixed, $\cHII$ is the only free parameter, which is constrained by fitting the predicted (simulated) $\Halpha$ LF to observational data. We fit the models to observed $\Halpha$ LF at two epochs. At $z\sim1$ we combine the NICMOS-based measurements of \citet{Yan_1999,Shim_2009} (35 sources at $0.7<z<1.4$) with the WFC3 measurements of \citet{Colbert_2013} (486 sources at $0.9<z<1.5$) and the recent JWST-based measurements of \citet{Pang_2026} (53 sources at $1.05<z<1.5$) into a single dataset of 21 points, covering $40.7\lesssim \log L_{\Halpha}\lesssim 43.3\ergss$. At $z\sim 1.5$ we use the narrow-band measurements of \citet{Sobral_2013}, comprising 13 points constructed from 346 sources selected either at $1.35<z<1.55$ or through the $z\sim 1.5$ B$z$K criterion, and covering $41.8\lesssim \log L_{\Halpha}\lesssim 43.0\ergss$. These LF are as-observed, i.e. not corrected for dust extinction, and we therefore compare them with the dust-extincted luminosities of the simulated ELGs.

The $z\sim 1$ data are compared to the $z=1$ snapshots of TNG100-1, TNG300-1 and Illustris-1, and the $z\sim 1.5$ data to the $z=1.5$ snapshot of TNG100-1 and the $z=1.47$ snapshot of Illustris-1. We apply our emission models on all subhalos which contain at least one gas cell, with the number counts given in Table \ref{tab:subhalo_counts}. For each target subhalo, we sample the dust-extincted $\Halpha$ luminosity for each model for an array of $\cHII$ values with logarithmic spacing $\Delta \log \cHII = 0.005$ such that both the $\Halpha$ luminosity output and the $\Halpha$ extinction of the galaxies are computed simultaneously. The results are binned in $\Delta \log L_{\Halpha}=0.25$ to obtain the model LF, which are then compared to the observed LF to sample the normalized $\chi^2$ of the models (the standard $\chi^2$ divided by the number of degrees of freedom, for a single free parameter). The $\chi^2$ is weighted solely by the published observational error bars of the data points, and no uncertainty on the simulated LF is included. We do not mask any bin, including bins at $\log L_{\Halpha}\gtrsim 42.8 \ergss$, which contain only $\sim O(10)$ ELGs and are therefore subject to significant sampling noise. A single ELG in a bin would correspond to $\log \phi = -5.53,\ -5.48\ \rm{Mpc^{-3}dex^{-1}}$ for TNG100-1 and Illustris-1, respectively. We verified that repeating the calculation with the LOS along the $x$ and $y$ axes of the simulation box, instead of the $z$ axis, changes the LF by less than $0.1$ dex for the same $\cHII$, including in these bins. Finally, we note that the best-fit $\cHII$ depends on the sputtering temperature threshold adopted in \S\ref{sec:dust}. Varying it between $10^{5-6}\Kel$ can shift $\cHII$ by a factor of a few, and $2\times10^5\Kel$ was found to yield $\cHII \approx 41.3$. The obtained best-fit $\chi^2$, however, varies strongly only for Illustris.

%----------------
\begin{figure*}[htp]
    \centering
    \includegraphics[width=0.7\textwidth]{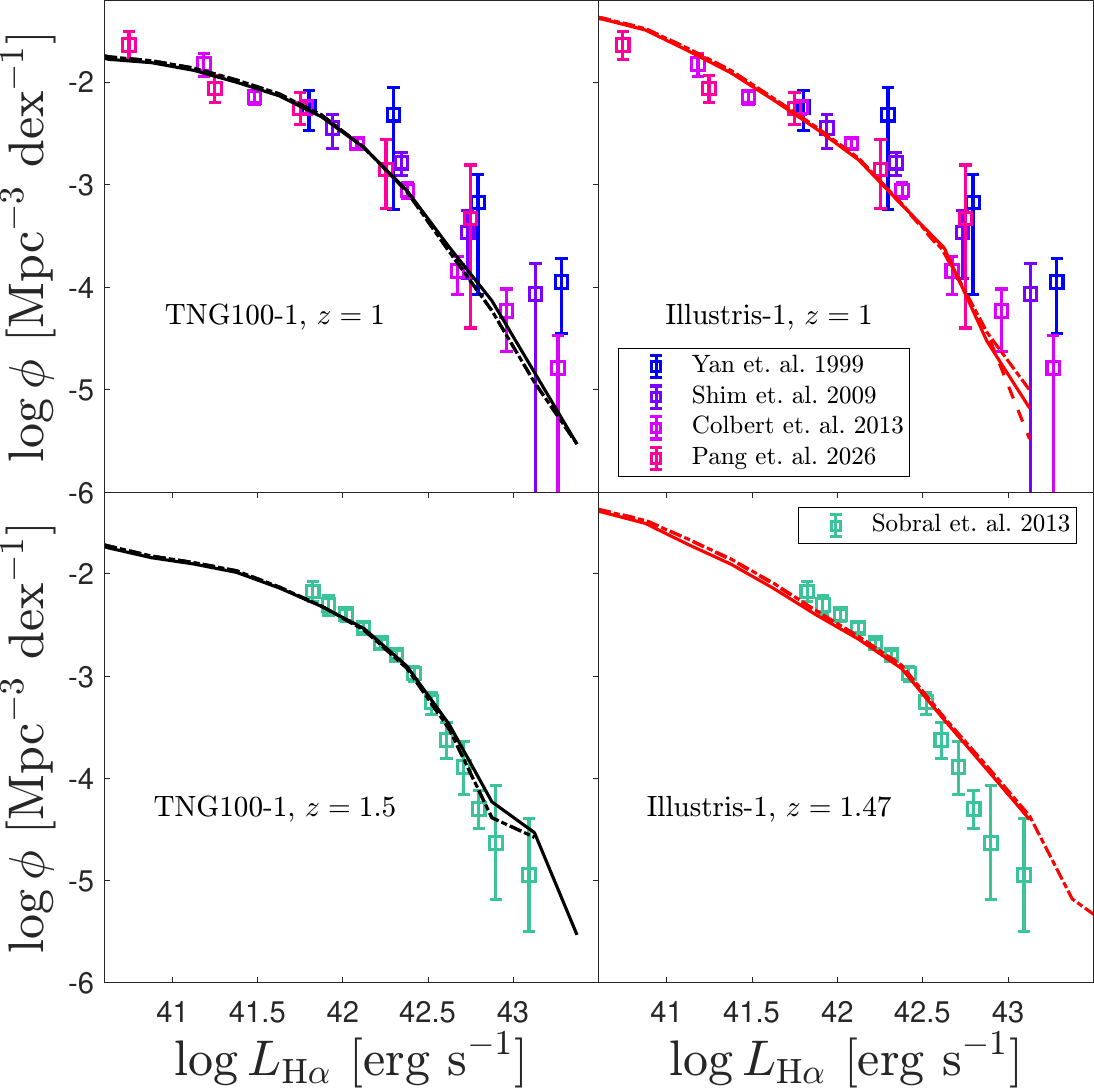}
    \caption{Best-fit LF for each model in the two simulations, shown in separate columns. Solid, dashed, and dot-dashed lines correspond to the HII, HII+Coll, and HII+Coll+AGN models, respectively. These are compared to the observed $\Halpha$ LF bracketing the simulation redshift.}
    \label{fig:LF_best_fits}
\end{figure*}
%--------------

The best-fit $\Halpha$ LF are shown in Fig.~\ref{fig:LF_best_fits} along with the data, while Table \ref{tab:best_fit_properties} summarizes the best-fit values of $\cHII$ and normalized $\chi^2$. The number of degrees of freedom is 20 ($z=1$) and 12 ($z\sim 1.5$).
All the predictions extracted from the low-feedback TNG simulations are consistent with the data at redshift $z\sim 1$ and 1.5, with a best-fit value of $\cHII$ decreasing when the collisional components are included in the fit. By contrast, the best-fit models extracted from the (high) feedback Illustris simulation show a statistically significant discrepancy with the data, with $p$-values $\simeq 0.005$ ($z=1$) and $\simeq 0.02$ ($z=1.47$). 
Including AGN photo-excitation has no effect on the best-fit LF compared to the HII+Coll model, except for TNG300-1 where the AGN-induced contribution changes the inferred $\cHII$ by $\sim 1\%$ and yields a small reduction in $\chi^2$.

%----------------
\begin{figure*}[htp]
    \centering
    \includegraphics[width=0.7\textwidth]{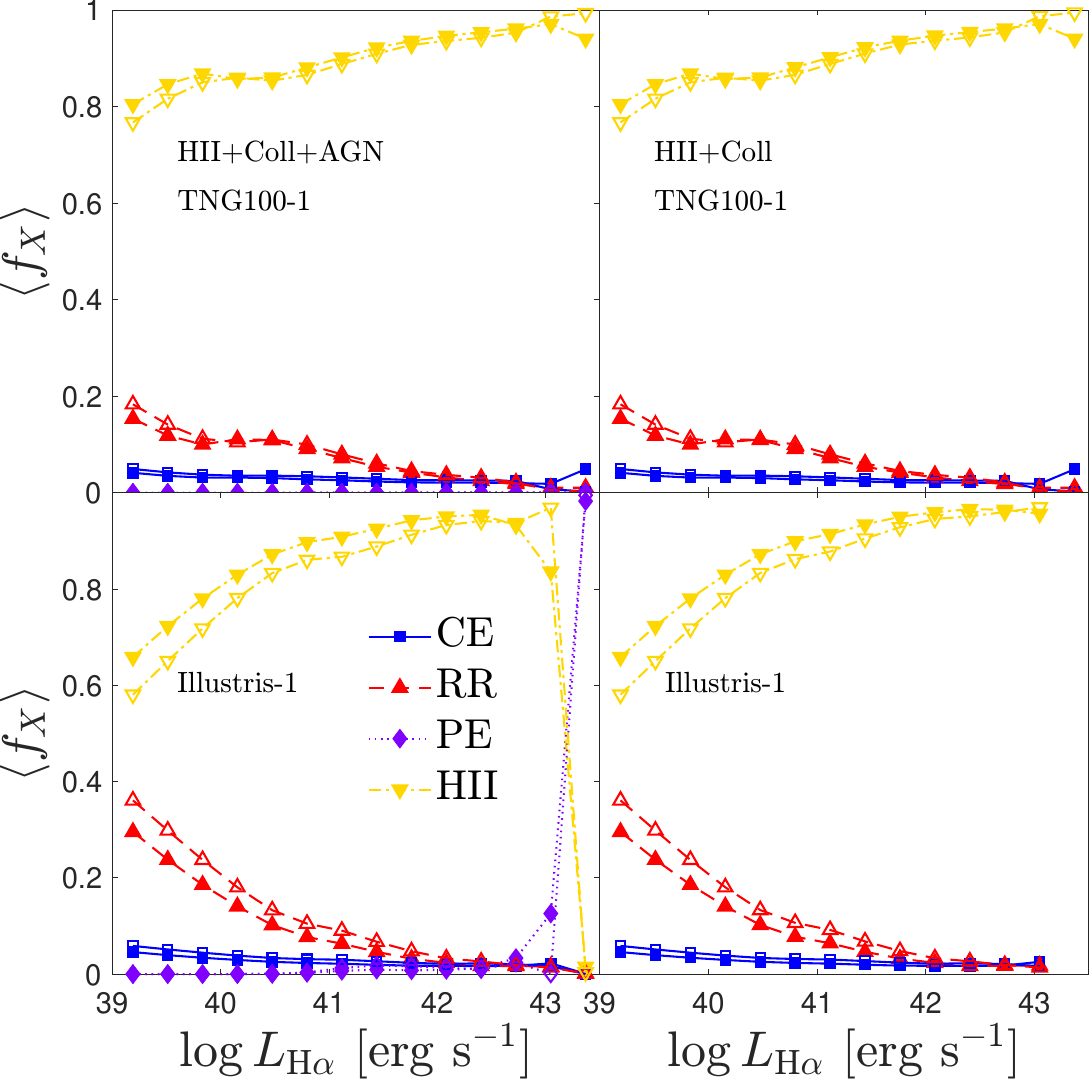}
    \caption{Average contribution fractions $\langle f_X\rangle\equiv \left\langle L_{\Halpha}^X/L_{\Halpha} \right\rangle$ for the different sources of $\Halpha$ emission. Results are shown for the 
    best-fit HII+coll+AGN model (left panels) and HII+Coll model (right panels) as a function of the total $\Halpha$ luminosity of galaxies. The upper row is for TNG100-1, whereas the lower one is for Illustris-1. Filled markers indicate $\langle f_X\rangle$ at $z=1$, while the empty markers show $\langle f_X\rangle$ at $z=1.5$ (TNG) or $z=1.47$ (Illustris). }
    \label{fig:cont_frac_lumi_bins}
\end{figure*}
%--------------

In Fig.~\ref{fig:cont_frac_lumi_bins}, we present, for the best-fit models of each simulation, the mean fractional contributions of the different $\Halpha$ emission components described in Section~\ref{sec:emission_model}, as a function of total $\Halpha$ luminosity. In the HII+Coll model, the $\Halpha$ emission at redshift $z\sim 1 - 1.5$ is dominated by HII regions, with an average contribution of $\langle f_{\rm HII} \rangle \approx 85\% - 95\%$, while collisional excitation (CE) and radiative recombination (RR) together account for the remaining $\sim 5\% - 15\%$.
The Illustris simulation predicts a larger contribution from the diffuse ISM than TNG at the faint end of the LF.
These relative fractions do not change appreciably for TNG when photo-excitations from AGN are included, but in Illustris the AGN-induced $\Halpha$ emission becomes dominant above $L_{\Halpha} \approx 10^{43}\ergss$. We caution that these bins contain few galaxies, so this outcome is subject to sampling noise.
Unlike the TNG simulations, the Illustris snapshots at $z=1$ and $1.47$ suggest a mild redshift evolution: at $z = 1$, the relative contributions of RR is slightly reduced compared to $z = 1.47$, with a corresponding increase in the HII contribution.

%----------------
\begin{figure}
    \centering
    \includegraphics[width=0.42\textwidth]{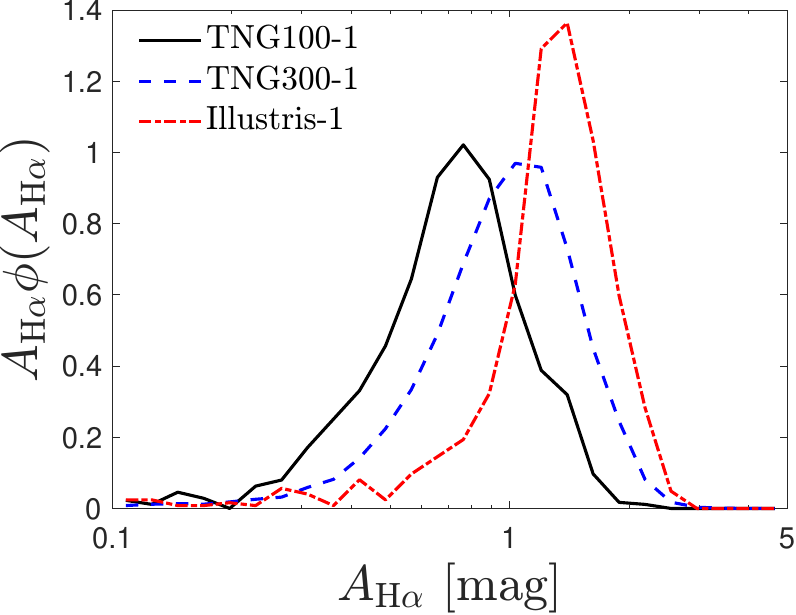}
    \caption{PDFs of the galactic $\Halpha$ extinction magnitudes ($A_{\Halpha}\equiv 1.086\bar{\tau}_{\Halpha}$) for the resulting best fits of the HII+Coll+AGN model for different simulations at $z=1$ and for galaxies with $L_{\Halpha}>10^{42} \ergss$. The PDFs include both inclination and intrinsic scatter.}
    \label{fig:ext_dist}
\end{figure}
%--------------

\subsection{Implied $\Halpha$ extinction}

Once the parameter $\cHII$ has been fitted to the data, the intrinsic $\Halpha$ luminosity of each galaxy is fully determined, allowing us to evaluate the $\Halpha$ extinction implied by the model.
For this purpose, we define an effective, galaxy-integrated $\Halpha$ optical depth $\bar{\tau}_{\Halpha}$ by equating the total observed luminosity, $L_{\Halpha}$, to that of a uniform foreground screen,
\begin{equation}
    L_{\Halpha} \equiv L_{\Halpha}^{\text{intr}}e^{-\bar{\tau}_{\Halpha}} \;,
\end{equation}
or
\begin{equation}
\bar{\tau}_{\Halpha}
=
-\ln\!\left(
\frac{L_{\Halpha}}
{L_{\Halpha}^{\text{intr}}}
\right).
\end{equation}
Fig.~\ref{fig:ext_dist} shows the probability distribution function (PDF) of the resulting $\Halpha$ extinction magnitudes ($A_{\Halpha}\equiv 1.086\bar\tau_{\Halpha}$) for bright, $z=1$ galaxies with luminosity $L_{\Halpha}>10^{42}\ergss$, for the HII+Coll+AGN model. Note that these PDFs combine the scatter due to the inclination of galaxies with respect to the LOS with the intrinsic scatter of their dust content. Their mean is insensitive to the choice of LOS, but the width of the distribution is not purely intrinsic. The two TNG simulations yield mean extinction magnitudes of $\langle A_{\Halpha} \rangle=0.73$ mag (TNG100-1) and $\langle A_{\Halpha} \rangle=0.99$ mag (TNG300-1), the latter showing a non-negligible impact on the extinction magnitudes due to coarser resolution. Illustris has the highest mean extinction magnitude of $\langle A_{\Halpha} \rangle=1.25$ mag. These values are broadly consistent with models calibrated on local galaxies \citep[e.g.][]{Charlot_2000,chevallard/etal:2013} as well as high redshift surveys \citep{kashino/etal:2013, ibar/etal:2013}, which is partly by construction, due to the chosen sputtering temperature threshold (Sec.~\ref{sec:dust}). Since TNG100-1 and Illustris-1 have a similar resolution, we conclude that the feedback prescription is the dominant source in the differences between $A_{\Halpha}$, with resolution playing a secondary role.

\subsection{Theoretical prediction for $\cHII$}

To interpret the values of $\cHII$ inferred from our fits, we derive a theoretical estimate based on the ionizing output of a continuously star-forming stellar population. First, we calculate the ionizing photon emission rate of a single star and its associated $\Halpha$ luminosity due to recombinations in the surrounding gas. We then estimate the population of ionizing stars using the star formation rate, the stellar initial mass function, and their finite lifetimes, and integrate over the stellar population to obtain $\cHII$.

We begin by estimating the production rate of hydrogen-ionizing photons, $Q_{\text{H}}$ (in $\rm{s^{-1}}$), emitted by a single star. 
Approximating the stellar spectrum as a blackbody, we get
\begin{align}
    & Q_{\text{H}} = \frac{8\pi^2 h R_*^2 c^2}{(hc)^4} \int_{I_\infty}^\infty \frac{E^2}{e^{E/kT_*}-1}dE \\ 
    &\approx \frac{8\pi^2 h R_*^2 c^2}{(hc)^4}(kT_*)^3e^{-\frac{I_{\infty}}{kT_*}}\left(2+2\frac{I_\infty}{kT_*}+\left(\frac{I_\infty}{kT_*}\right)^2\right) \nonumber \;,
\end{align}
where $k$ is the Boltzmann constant, $h$ is the Planck constant, $c$ is the speed of light, $I_\infty=13.6 \eV$ is the H\,I ionization energy (Rydberg energy) and $T_*$, $R_*$ are the effective black-body temperature and radius of a star with mass $m_*$. They satisfy the relations \citep{Demircan1991}
\begin{align}
    &T_* = (5785\Kel)\times \left(\frac{m_*}{M_\odot}\right)^{0.475} \nonumber \\ & R_* = (6.96\times 10^{10}\ \text{cm})\times \left(\frac{m_*}{M_\odot}\right)^{0.8} \;.
\end{align}
Assuming every ionizing photon is absorbed by H\,I and ultimately produces a recombination, the $\Halpha$ luminosity associated with a single star is
\begin{equation}
    L_{\Halpha,*} = Q_{\text{H}} E_{\Halpha} \mathcal{P}_{\Halpha} \;,
\end{equation}
where $E_{\Halpha}=1.89\eV$ is the $\Halpha$ photon energy and $\mathcal{P}_{\Halpha} \approx 0.15$ is the probability that a $\Halpha$ is produced in a H recombination cascade. 
The $\Halpha$ luminosity arising from H II regions can be written as an integral over the stellar population,
\begin{equation}
    L_{\Halpha}^{\text{HII}} = \int  L_{\Halpha,*}\left(\frac{dN_*}{d m_*}\right)dm_* \;,
\end{equation}
where $dN_*/dm_*$ is the population density of stars with mass $m_*$. For a continuously star-forming region, the stellar population is determined by the competition between stellar birth and death. We denote $\phi_*(m_*)$ the stellar mass fraction distribution of newly created stars, defined as
\begin{equation}
    \phi_*= \frac{m_*\xi(m_*)}{\int m_*\xi(m_*) dm_*}\;,
\end{equation}
where $\xi(m_*)$ is the Kroupa initial mass function \citep{Kroupa_2001}. In the following, we restrict the computation to the ionizing stellar population $m_*\in(15,150)M_\odot$, and normalize $\phi_*\propto m_*^{-1.3}$ over this range. The present population density of stars with mass $m_*$ depends on the star formation rate through 
\begin{align}
    & \frac{dN_*}{dm_*}=\frac{\phi_*}{m_*}\int_{t-t_*}^t \text{SFR}(t')dt' \\ & \approx \frac{\phi_*}{m_*} t_*\text{SFR} \;, \nonumber
\end{align}
where only stars formed during the previous ionizing lifetime $t_*(m_*)$ remain capable of contributing to the present ionizing radiation field. Our approximation also assumes that the SFR remains constant over the short lifetimes of massive stars ($t_*\lesssim 10\ \text{Myr}$ for $m_*\gtrsim 15 M_\odot$). The H II region luminosity therefore becomes
\begin{equation}
    L_{\Halpha}^{\text{HII}} = \cHII\text{SFR} \;,
\end{equation}
with
\begin{equation}
    \cHII = \int L_{\Halpha,*}\frac{t_*}{m_*}\phi_*dm_* = \left\langle \frac{L_{\Halpha,*} t_*}{m_*}\right\rangle_{\phi_*} \;. 
\end{equation}
$\cHII$ may be interpreted as the total energy emitted by the H II regions in $\Halpha$ photons over the star's lifetime, per unit of stellar mass formed and averaged over $\phi_*$. 
Adopting a mass-dependent lifetime
\begin{equation}
    t_*=10\ \text{Myr} \times \left(\frac{m_*}{15M_\odot}\right)^{-\alpha_*} \;,
\end{equation}
with a constant slope $\alpha_*=2$ appropriate for the main sequence \citep[e.g.][]{Romano_2004}, we arrive at
\begin{equation}
\label{eq:cHII_theory}
    \cHII \approx 10^{41.50} \ergss \rm{yr M_\odot ^{-1}} \;,
\end{equation}
which is higher by factor $\sim 2$ compared to the best-fit values of $\cHII$ inferred from the TNG100-1 and Illustris LF data. 
Note, however, that this estimate is sensitive to the behavior of $\phi_*$ and $t_*$. 
Taking a steeper slope of $\alpha_*=2.5$ results in a value of $\cHII$ 10 - 20\% lower than the TNG best fits. 
On the other hand, taking $\alpha_*=0$ (constant $t_*=10\ \text{Myr}$) results in an unrealistically large $\cHII=10^{43.0} \ergss \rm{yr M_\odot ^{-1}}$. 
Moreover, Eq.~\eqref{eq:cHII_theory} approximates stellar spectra as blackbodies, whereas the UV flux of real stars show a deficit compared to that of a blackbody owing to line blanketing. This effect would lower the value of $\cHII$ and may account for part of the difference.

To conclude, our $\cHII$ determines the intrinsic $\Halpha$ luminosity of H\,II regions. Since extinction is applied separately in our model, the fitted $\cHII$ should be compared directly with $\Halpha$--SFR relations for extinction-corrected luminosities. These give $\log\cHII\approx 41.1$ for a \citet{Salpeter_1955_IMF} IMF \citep{Kennicutt_1983, Dominguez_Sanchez_2012}, $41.27$ for a Kroupa IMF \citep{kennicutt/evans:2012} and $\approx 41.2$ for a \citet{Chabrier_2003_IMF} IMF \citep{hirschmann_2023}. TNG and Illustris adopt a Chabrier IMF, so the Chabrier and Kroupa values are the most directly comparable; the Salpeter values are lower because a Salpeter IMF yields a larger SFR for a given luminosity. Our best-fit values for TNG100-1 and Illustris-1 ($\log\cHII=41.16$--$41.34$) agree with the Chabrier and Kroupa values to within $\sim 0.1$ dex, whereas the theoretical estimate of Eq.~\eqref{eq:cHII_theory} is higher by a factor of $\sim2$, an offset within the uncertainties of that estimate discussed above.

%----------------------------------------------------------------------------%
\section{Discussion}
\label{sec:discussion}
%----------------------------------------------------------------------------%

The $\Halpha$ emission line model used here adopts a treatment of H II regions and dust attenuation different from \cite{Rapoport_2025}.
In \cite{Rapoport_2025}, $\cHII$ was fixed to the value of $\cHII=10^{40.8}\ergss \rm{yr M_\odot ^{-1}}$, while the dust attenuation was uniform across all galaxies, 
with values in the range $0.3\lesssim A_{\Halpha}\lesssim0.85$ mag providing good agreement with the observed LF. 
The contribution from CE and RR was also larger owing to the different treatment of star-forming cells. 
In the present work, dust attenuation is computed on a galaxy-by-galaxy basis using the dust-to-gas ratio prescription of \cite{Bialy_Sternberg_2019}, calibrated using galaxies in the local Universe, and including the destruction of dust grains by thermal sputtering in hot gas,
while $\cHII$ is treated as a free parameter to account for the unresolved star-forming regions. The \citet{Bialy_Sternberg_2019} prescription was calibrated on local-Universe galaxies, and its applicability at $z\sim 1-1.5$ is subject to uncertainties in the evolution of dust properties and its distribution within galaxies; nevertheless, we consider it broadly applicable within these uncertainties, given the lack of a directly calibrated alternative at these redshifts. The inclusion of dust sputtering is essential: without it, Illustris fails to reproduce the shape of the observed LF at $z\sim 1$, and in both simulations heightened values of $\cHII$ are obtained.
Given these modeling choices, we now turn to the resulting differences in the physical picture. In the present model, the mean extinction implied for bright galaxies at $z=1$ is in the range $\langle A_{\Halpha}\rangle=0.73$--$1.25$ mag, comparable or somewhat above to the assumed values in \citet{Rapoport_2025}, while the fitted $\cHII$ is larger.
This implies relatively smaller contributions of CE + RR to the $\Halpha$ emission. This result is consistent with \citet{Tacchella_2022}, who model $\Halpha$ emission in the Milky Way and the Large Magellanic Cloud and find a similarly small ($5\%-10\%$) contribution from collisional processes. However, narrow-band surveys of local galaxies indicate that emission from the diffuse ionized gas (DIG) may contribute as much as $70\%$ to the total $\Halpha$ emission \citep{Zurita_2000}. 
Separately, we verified that our results for the AGN-induced $\Halpha$ contributions are robust to variations in the slope $\alpha_\text{AGN}$ of the AGN spectral energy distribution, since AGN feedback is significant at the bright end only. 

Our analysis shows that the Illustris simulation provides a poorer agreement with the data at $z=1$. 
This discrepancy persists at $z\sim 1.5$, where Illustris fails to reproduce the $z\sim 1.5$ observations both at the faint and bright ends. 
The SFR function (SFRF) could provide an interesting complementary test. For example, \cite{gruppioni/etal:2015} measured the SFRF out to $z\sim 3$ using far-IR and UV observations. 
A direct comparison of these measurements with the SFRF predicted by Illustris and IllustrisTNG would provide another way of testing whether their different feedback prescriptions 
reproduce the evolution of the SFR.

The resulting preference for the low-feedback IllustrisTNG is at odds with recent DESI - ACT measurements of the stacked kinetic Sunyaev-Zel'dovich (kSZ) temperature decrement \citep{tSZ1972,kSZ1980}, 
which reconstruct averaged radial gas profiles and can constrain the impact of baryonic feedback.
At redshift $z\lesssim 1$, the observed kSZ signal is better reproduced by the high-feedback Illustris simulation than the low-feedback IllustrisTNG simulation \citep{Hadzhyska_2025,Guachalla_2025,bigwood/etal:2025}. A preference for stronger feedback is echoed by independent thermal Sunyaev-Zel'dovich (tSZ) measurements around similar galaxy samples \citep{Siegel_2026}.

To understand this tension, note that the kSZ signal traces the integrated momentum of free electrons along the line of sight, which are predominantly found in the diffuse circumgalactic medium rather than within galaxies themselves. The $\Halpha$ emission, by contrast, is sourced primarily by cooler, star-forming gas, with a smaller contribution from collisional excitation and recombination in the diffuse ISM. Therefore, the kSZ signal is sensitive to the extended distribution of ionized gas around galaxies, while $\Halpha$ traces denser and cooler gas phases within them. The fact that Illustris simultaneously reproduces the stronger kSZ signal but overpredicts the observed $\Halpha$ LF at the faint end may therefore indicate 
that its feedback model produces an excess of gas across multiple phases, rather than a simple redistribution of baryons to larger radii. 
By contrast, the low-feedback IllustrisTNG simulation yields star formation rates and neutral hydrogen abundances that produce galaxy $\Halpha$ luminosities consistent with the observations,
but it underpredicts the large-scale free-electron distribution inferred from kSZ measurements. 
This may point to remaining deficiencies in how current simulations model the multiphase ISM and the impact of feedback on the circumgalactic medium. 

Summarizing, the combination of stacked kSZ and $\Halpha$ LF is a powerful probe of the interplay between feedback and star formation because kSZ measurements are sensitive to the amount and spatial distribution of ionized gas 
while the $\Halpha$ LF depends on the amount of gas in the warm ISM phase and the efficiency with which gas cools and forms stars.

%----------------------------------------------------------------------------%
\section{Conclusions}
\label{sec:conclusions}
%----------------------------------------------------------------------------%

We have explored the impact of baryon feedback on the $\Halpha$ luminosity function (LF) at redshift $z\sim 1$ using the high-feedback Illustris and low-feedback IllustrisTNG simulations. 
For this purpose, we have improved the modelling of the galactic $\Halpha$ emission developed in \citet{Rapoport_2025, Rapoport_2026_environment} by adding, on a galaxy-by-galaxy basis, an explicit directional computation of $\Halpha$ extinction and a physically informed model of the mean free path of Lyman-exciting photons, which affects the AGN-induced $\Halpha$ emission. Using this framework, we consider three models of increasing complexity: an HII model including only H II-region emission (with the collisional/recombination component switched off), an HII+Coll model that adds collisional excitation and recombination in the diffuse ISM, and an HII+Coll+AGN model that further includes excitations by photons originating in AGN. The models have a single main free parameter, the conversion constant $\cHII$ relating star formation rate to H II-region $\Halpha$ luminosity ($L_{\Halpha}^{\text{HII}}=\cHII \text{SFR}$), which we calibrate against LF observations at $z\sim 1-1.5$.

Our results can be summarized as follows:
\begin{itemize}
    \item For the key parameter $\cHII\equiv L_{\Halpha}^{\text{HII}}/\text{SFR}$, the best-fit values of TNG100-1 and Illustris-1 lie in the range $\cHII= 10^{41.16}-10^{41.34} \ergss \text{yr}M_\odot^{-1}$ across all redshifts and models, consistent with simulations and observations calibrated on Kroupa or Chabrier IMF. TNG300-1 has an irregularly high value of $\cHII\approx 10^{41.6} \ergss \text{yr}M_\odot^{-1}$ which we attribute to its lower resolution. In both TNG100-1 and Illustris-1, a slightly lower value of $\cHII$ is obtained at $z\sim 1.5$. 

    \item Our dust modelling implies a mean extinction of $\Halpha$ in the source rest frame of $\langle A_{\Halpha} \rangle\approx 0.7-1.3$ mag for all simulations, consistent with high-redshift surveys and models calibrated on data. Accounting for the destruction of dust grains by thermal sputtering in hot gas is essential to reproduce the Illustris LF. The adopted sputtering threshold affects the obtained best-fit $\cHII$. 
    
    \item Likewise, the absorption and attenuation of photo-exciting photons leaving galactic centers in AGN-host galaxies depends strongly on the galaxy formation model. In particular, the attenuation length of Lyman-series photons with energy $E\geq E_{{\rm Ly}\beta}$ is shorter in the Illustris simulation.
    \item Fitting $\cHII$ to the observed $\Halpha$ LF implies a 5\% - 15\% contribution from collisional processes in the diffuse gas in the warm ISM to the total $\Halpha$ emission for galaxies with $L_{\Halpha}\geq 10^{41}\ergss$. The best-fit HII and HII+Coll models yield very similar $\Halpha$ LF. In Illustris, when the photo-excitation of diffuse gas by AGN radiation is taken into account, it becomes the dominant source of $\Halpha$ emission for the bright $\Halpha$ emitters with $L\gtrsim  10^{43}\ergss$.
    
    \item Our best-fit values of $\cHII$ are consistent (to within $\sim 0.1$ dex) with empirical and simulated $\Halpha$--SFR relations for extinction-corrected luminosities that assume a Kroupa or Chabrier IMF. We find that $\cHII$ is constrained mainly by the faint end of the LF, where observational uncertainties are smallest. 
    
    \item The low-feedback IllustrisTNG simulations provide a reasonable fit to the $\Halpha$ LF for the redshifts and models considered here. By contrast, the high-feedback Illustris 
    simulations show statistically significant departures from the data at $z=1$ and 1.5 both at the bright and faint end of the LF. The statistical significance is driven primarily by the 
    faint end, which carries the most constraining power because of its small observational uncertainties.
    
    \item Our findings are in tension with recent analyses of stacked kSZ measurements at $z\lesssim 1$ \citep{Hadzhyska_2025}, which suggest that the high-feedback Illustris simulations are a better fit to the kSZ data. 
\end{itemize}
Using \textit{Euclid} and NGRST data to improve measurements of the faint-end $\Halpha$ LF could provide much stronger constraints on physically motivated models of $\Halpha$ emission. Combined with kSZ measurements, they could constrain baryon feedback in high-redshift galaxies. 
We expect the modeling framework developed in this work to provide a useful basis for interpreting these data and for improving the physical connection between galaxy formation models and observations.

%----------------------------------------------------------------------------%
\section{Acknowledgements}
V.D. thank Yakov Faerman, Adi Nusser and Jonathan Stern for interesting discussions.
We thank Pierluigi Monaco and Emmanuel Schaan for helpful comments on an earlier draft.
E.B. acknowledges support by The Israel Science Foundation (grant No.~2617/25). S.B. acknowledges support from the ISF grant number 2071540 and the Alon Fellowship prize for junior faculty.

\appendix

%----------------------------------------------------------------------------%
\section{Modelling intrinsic $\Halpha$ emission}
\label{sec:appendix}
%----------------------------------------------------------------------------%

\subsection{Cell Properties}

In \citet{Rapoport_2025}, the intrinsic $\Halpha$ emission from galaxies is modelled by summing the hydrogen line emission computed for individual gas cells within each subhalo hosting a galaxy.
Each gas cell is described by its hydrodynamic properties $(m_{\rm gas}, \rho_{\rm gas}, X_{\rm H}, x_e, x_{\rm H\,I}, u, {\rm SFR}, f_{\rm AGN})$, corresponding to: the total gas mass, the gas density, the hydrogen fractional mass, the electron and neutral hydrogen abundance ratios, the internal energy per unit mass, the instantaneous star formation rate and the local AGN flux, respectively. From these properties we compute for each cell the volume $V_{\rm gas}$, the neutral and total hydrogen number densities $n_{\rm H\,I}, n_{\rm H}$ and a temperature $T$ assuming an ideal gas equation of state. These quantities are used to compute the $\Halpha$ output of cells in simulated galaxies.

\subsection{Treatment of star-forming cells}

Gas above the star-formation density threshold is modeled using the effective two-phase subgrid ISM model of \citet{Springel_2003}, in which cold clouds are embedded in a hot ambient medium. For these star-forming cells, we compute the hydrodynamic properties of the hot phase as detailed below and apply the hydrogen level population model only to this component.

The internal energy per unit mass provided by the simulation is the mass-weighted average of the hot ($u_h$) and cold ($u_c$) phases,
\begin{equation}
\label{eq:u_h}
    u = x_h u_h + (1-x_h)u_c \approx x_h u_h,
\end{equation}
where $x_h$ is the hot-phase mass fraction. Since the cold phase is assumed to have a much lower temperature than the hot phase ($T_c\ll 10^4\Kel$), its contribution to the internal energy is negligible, i.e. $u_c\ll u_h,$ yielding the approximation in Eq.~(\ref{eq:u_h}). We determine $x_h$ for each star-forming cell by interpolating the equilibrium solution of \citet{Springel_2003} as a function of the cell baryonic density. Inverting Eq.~(\ref{eq:u_h}) then yields the hot-phase internal energy, from which we compute the hot-phase temperature, $T_h$, assuming a mean molecular weight of $\mu_h=0.59$, appropriate for fully ionized primordial gas.

The hydrogen number density in the hot phase is given by
\begin{equation}
    n_{\mathrm{H},h}=\frac{X_{\mathrm{H}}\rho_{\mathrm{gas}}x_h}{m_p},
\end{equation}
where $m_p$ is the proton mass. Since the hot phase is assumed to be nearly fully ionized, we approximate the electron number density as $n_{e,h}=n_{\mathrm{H},h}$. The residual neutral hydrogen density is then obtained by imposing collisional ionization equilibrium,
\begin{equation}
    n_{\mathrm{H\,I},h}
    =
    \frac{\alpha(T_h)}
    {\alpha(T_h)+C(T_h)}
    \,n_{\mathrm{H},h},
\end{equation}
where $\alpha(T_h)$ and $C(T_h)$ are the temperature-dependent recombination and collisional ionization rate coefficients, respectively. In the hot phase of star forming cells, typically $n_{\mathrm{H\,I}}/n_{\mathrm{H}}\approx 10^{-5.5\pm 1.5}$. 

\subsection{Hydrogen level population model for the diffuse ISM}

We compute $\Halpha$ emission from the diffuse ISM by solving for the steady-state hydrogen level populations in each gas cell. 
The populations $n_i$ of bound levels and the continuum satisfy
\begin{equation}
    \frac{dn_i}{dt}=\sum_j Q_{ij}n_i =0 \;.
\end{equation}
In practice, we restrict our calculation to the hydrogen levels with principal quantum numbers $1\leq n \leq 5$ and the ionized state. 
The transition matrix $Q_{ij}$ includes radiative decay, collisional excitation (CE) and recombination (RR) and photo-excitation due to AGN flux (AGN). 
The $\Halpha$ photon production rate per hydrogen atom is obtained from the level populations as
\begin{equation}
    R_{\Halpha}= \sum_{i\in n=3}\sum_{j\in n=2}n_iA_{ij} \;,
\end{equation}
where $A_{ij}$ are the Einstein coefficients for the $i\to j$ relaxation, and the sum is over fine-structure levels with principal quantum numbers $n=3$ and $n=2$. Due to the low-density conditions of the ISM, we can solve the contribution for each process separately to obtain $R_{\Halpha}^{\rm CE}$, $R_{\Halpha}^{\rm RR}$, $R_{\Halpha}^{\rm AGN}$ for the aforementioned processes. 
\subsection{Total emission model}
The total intrinsic $\Halpha$ luminosity of a galaxy is obtained by summing contributions from all gas cells and adding a star-formation–calibrated H II component:
\begin{equation}
    L_{\Halpha}^{\text{intr}} = E_{\Halpha}\sum_{\text{gas cells}}\left(V_{\text{gas}}n_{\rm H\,I}\left[R_{\Halpha}^{\text{CE}}+R_{\Halpha}^{\text{AGN}}\right]+V_{\rm gas}n_{\rm H}R_{\Halpha}^{\rm RR}\right) + L_{\Halpha}^{\rm HII} \;,
\end{equation}
where $E_{\Halpha}\approx 1.89\eV$ is the $\Halpha$ photon energy. Since H II regions are unresolved, we model their emission separately as
\begin{equation}
    L_{\Halpha}^{\rm HII}= \cHII\, \rm{SFR_{gal}} \;,
\end{equation}
where $\rm{SFR_{gal}}$ is the total galactic SFR. 
\bibliography{references}

\end{document}